\documentclass{aa}
\usepackage{natbib}
\usepackage[varg]{txfonts}
\usepackage{booktabs}
\usepackage[colorlinks=true,linkcolor=blue,citecolor=blue,urlcolor=blue]{hyperref}
\bibpunct{(}{)}{;}{a}{}{,}
\usepackage{color}

\begin{document}

\title{Iron-rich solar particle events measured by SOHO/ERNE during two solar cycles}
\titlerunning{Iron-rich solar particle events during two solar cycles}
\author{O.~Raukunen \and E.~Valtonen \and R.~Vainio}
\authorrunning{O. Raukunen et al.}
\institute{Department of Physics and Astronomy, University of Turku, 20014 Finland \\
e-mail: \texttt{osku.raukunen@utu.fi}}
\date{Received <date> / Accepted <date>}
\abstract
{}
{We study the differences in the heavy ion composition of solar energetic particle (SEP) events between solar cycles 23 and 24.}
{We have surveyed the SOHO/ERNE heavy ion data from the beginning of solar cycle 23 until the end of June 2015, that is, well into the declining phase of cycle 24. We used this long observation period to study the properties of heavy ions (from C to Fe) and to compare the two solar cycles in this respect. We surveyed the data for SEP events with enhancements in the Fe/C and Fe/O intensity ratios in the energy range 5--15 MeV per nucleon, and associated the events with solar flare and coronal mass ejections (CME) when possible. We studied the properties of heavy ions in these events and compared the average relative abundances of heavy ions between the two solar cycles.}
{We found that fewer days had C and O intensities higher than \textasciitilde 10 $^{-3}$~cm$^{-2}$sr$^{-1}$s$^{-1}$(MeVn$^{-1}$)$^{-1}$ during solar cycle 24 than during cycle 23. For Fe this difference was clear even at lower intensities. We also found that fewer days had Fe/(C+O) $>0.183$ during cycle 24. We identified 86 SEP events with at least one Fe-rich day, 65 of which occurred during cycle 23 and only 21 during cycle 24. We found that impulsive events have been almost completely absent during cycle 24. Mean abundances of heavy ions in the events were found to be significantly lower during cycle 24 than in cycle 23. Our results reflect the reduced solar activity in cycle 24 and indicate lower efficiency of particle acceleration processes for both gradual and impulsive SEP events in cycle 24.}
{}
\keywords{Sun: abundances - Sun: activity - Sun: particle emission}

\maketitle

\section{Introduction}\label{sect_intro}
The earliest observations of solar energetic particle (SEP) events were obtained with ionization chambers \citep{forbush1946a}. These events, known today as ground-level enhancement (GLE) events, were presumed to be caused by solar flares. In a comprehensive review of radio observations, \citet{wild1963a} suggested that the fast-drift type III radio bursts were produced by flare-accelerated outward streaming electrons, whereas slow-drift type II bursts were produced by electrons accelerated by shock waves that might also accelerate protons. This idea of two different physical mechanisms of SEP acceleration that contribute to the two classes of solar particle events, namely impulsive and gradual events, has been widely accepted \citep[e.g.,][]{reames2013a}. In the two-class paradigm, particle acceleration in impulsive flares has been associated with resonant wave-particle interactions \citep[e.g.,][]{fisk1978a,temerin1992a,miller1993a,zhang1995a,paesold2003a}, stochastic acceleration by plasma waves or turbulence \citep[e.g.,][]{mobius1980a,mobius1982a,petrosian2004a,liu2004a,liu2006a}, or magnetic reconnection \citep[e.g.,][]{drake2009a}, but none of the theories have been able to provide a complete description of the observed properties of impulsive events. Gradual events, on the other hand, are commonly described by diffusive acceleration at shock waves driven by coronal mass ejections (CMEs) \citep[e.g.,][]{cane1995a,cliver1999a}. Impulsive events typically have durations of a few hours, low fluences, and compact spatial scales, compared to the duration of several days, higher fluences, and extensive spatial scales of gradual events \citep[e.g.,][]{kahler1992a,reames1999a,reames2013a}.

A key feature in distinguishing the two classes of solar particle events has been the differences in their energetic particle composition. In general, impulsive events are considered to be electron rich, to have \element[][3]{He}/\element[][4]{He} ratios enhanced by a factor of up to $10^{4}$ , and Fe/O ratios enhanced by up to a factor of 10 compared to coronal values \citep[e.g.,][]{reames1985a,reames1988a}. Gradual events, on the other hand, have an energetic particle composition similar to that of the corona or solar wind \citep[e.g.,][]{meyer1985a}. Early measurements of heavy ion charge states showed that the mean ionic charge states in impulsive events were significantly higher than in gradual events \citep[e.g.,][]{klecker1984a,luhn1987a}. More recent studies at extended energy ranges have revealed a more complex picture: ion charge states in impulsive events are highly energy dependent, with $Q_{Fe}$ increasing by as much as six charge units between \textasciitilde 0.1--0.5 MeV \citep{difabio2008a}. In gradual events the mean charge states at low energies are similar to those of the solar wind, but show high variability at higher energies \citep[e.g.,][and references therein]{klecker2006a}. 

In reality, the separation of the two classes is not clear: residual suprathermal ions from impulsive events may contribute to the seed material accelerated by CME-driven shock waves, resulting in intermediate values for \element[][3]{He}/\element[][4]{He} or heavy element abundances \citep[e.g.,][]{mason1999a,torsti2002a,kocharov2002a,kocharov2003a,tylka2005a}. As a different explanation for the intermediate abundances, \citet{cane2003a,cane2006a} have argued in favor of a direct flare component or a combination of direct flare particles and the shock acceleration of these particles during large SEP events.

The current solar cycle, cycle 24, has shown a considerably lower overall activity level than the previous cycle 23. This has also left clear imprints on the SEP events. For example, GLEs have shown a dramatic decrease in number from 16 in cycle 23 to only one in cycle 24 \citep{gopalswamy2013a}\footnote{In addition, one SEP event has led to a counting-rate increase of two neutron monitors at the south pole \citep{thakur2014a}, but this is not included in the official list of GLEs (\url{http://gle.oulu.fi/}) because the event did not produce statistically significant increases in other stations}. In this paper we compare the number and properties of iron-rich SEP events for the two solar cycles. A preliminary analysis was performed in \citet{raukunen2015a}. The structure of this paper is as follows: in Sect. \ref{sect_obs} we present an overview of the observations we used, in Sect. \ref{sect_daily_int} we study the daily heavy ion intensities and intensity ratios during the previous and the current solar cycle, in Sect. \ref{sect_event_sel} we select a list of SEP events that contain enhancements in Fe, in Sects. \ref{sect_flares} and \ref{sect_prop_heavy} we study the solar event associations and properties of heavy ions in the selected events, and in Sect. \ref{sect_concl} we summarize the results and present the conclusions of the study. Appendix \ref{sect_stattest} includes technical details about the fitting and statistical testing.

\section{Observations}\label{sect_obs}
\begin{figure}
\resizebox{\hsize}{!}{\includegraphics{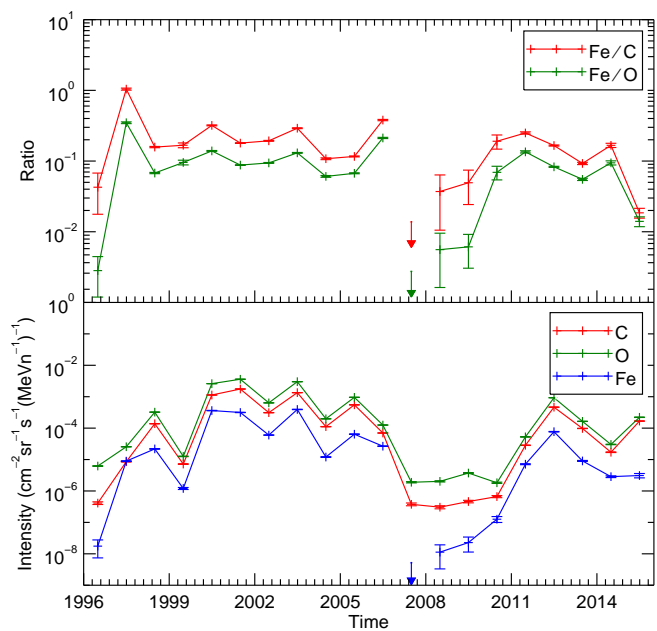}}
\caption{\emph{Bottom panel}: annual average intensities of 5--15 MeV~n$^{-1}$ C (red), O (green) and Fe (blue). \emph{Top panel}: annual average \mbox{5--15 MeV~n$^{-1}$} intensity ratios Fe/C (red) and Fe/O (green). The arrow symbols mark the one-count upper limits.} \label{intensity_vs_time}
\end{figure}
The particle observations were made with the Energetic and Relativistic Nuclei and Electron (ERNE) instrument onboard the Solar and Heliospheric Observatory (SOHO) \citep{torsti1995a,valtonen1997a}. ERNE consists of two particle detectors, the Low Energy Detector (LED) and High Energy Detector (HED). Our analysis was made in the energy range of 5--15 MeV~n$^{-1}$, which is measured by LED. It has a geometric factor of $0.260$--$0.915$ cm$^2$sr, depending on particle energy and species. As SOHO was launched on 2 December 1995 and is still operational, the observation period extends from the end of solar cycle 22 well into the current solar cycle 24. There have been some breaks in the observations, most notably from 25 June 1998 to 9 October 1998, from 21 December 1998 to 8 February 1999, and from 9 December 2012 to 8 February 2013. These, along with all the shorter breaks, have been taken into account in the following analysis.

As an example of the ERNE/LED long-term heavy ion data, the bottom panel of Fig. \ref{intensity_vs_time} shows the annual average intensities of 5--15 MeV~n$^{-1}$ C, O, and Fe for 1996--2015. The top panel shows the annual intensity ratios Fe/C and Fe/O for the same time period. The shape of the intensity profiles reflects the solar activity level. The values marked with arrows for 2007 are one-count upper limits because no Fe ions were detected during the year by the instrument in the 5--15 MeV~n$^{-1}$ energy range. The annual Fe/C and Fe/O ratios show fairly stable if slightly declining trends during the active years 1997--2006 and 2011--2014. During the years of low solar activity, the ratios Fe/C and Fe/O have average values of approximately $0.05$ and $0.005$, respectively, which are much lower than the value for Fe/O in the corona ($0.186$, measured at $1.4 \cdot 10^6$ K \citep{feldman2003a}), for example, or the values for Fe/O in the slow and fast solar wind ($0.120$ and $0.092$, measured at 1 keV and 2 keV, respectively \citep{vonsteiger2000a}).

We here compare the properties of heavy ions in solar cycles 23 and 24. The start times and the times of the solar activity maxima are presented in Table \ref{cycle_times}. We defined the start of the cycle as the middle of the month with the lowest monthly sunspot number during the corresponding minimum. Both November and December 2008 had the same sunspot number, therefore the start of cycle 24 was defined as 1 December 2008. Similarly, the time of maximum of the cycle was defined as the middle of the month with the highest monthly sunspot number. Cycles 23 and 24 were both double-peaked, and the maximum sunspot numbers were reached during the first peak of cycle 23 and the second peak of cycle 24. The sunspot data we used was the NOAA smoothed monthly international sunspot number \footnote{Available online at \\ \url{http://www.ngdc.noaa.gov/stp/space-weather/solar-data/solar-indices/sunspot-numbers/}}.

\begin{table}
\caption{Solar cycle data.}
\label{cycle_times}
\centering
\begin{tabular}{c c c c}
\hline \hline
Solar cycle & Min / Max & Date & SSN\tablefootmark{a} \\
\hline
23 & Min         & 1996--May--15  & $8.0$   \\
23 & Max         & 2000--Apr--15  & $120.8$ \\
24 & Min         & 2008--Dec--1   & $1.7$   \\
24 & Max         & 2014--Apr--15  & $81.9$  \\  
24 & End of obs. & 2015--June--30 & \ldots  \\
\hline
\end{tabular}
\tablefoot{\tablefoottext{a}{Smoothed monthly sunspot number from the NOAA international sunspot number listing.}}
\end{table}

\section{Results and discussion}\label{sect_results}
\begin{figure*}
\centering
\includegraphics[width=17cm]{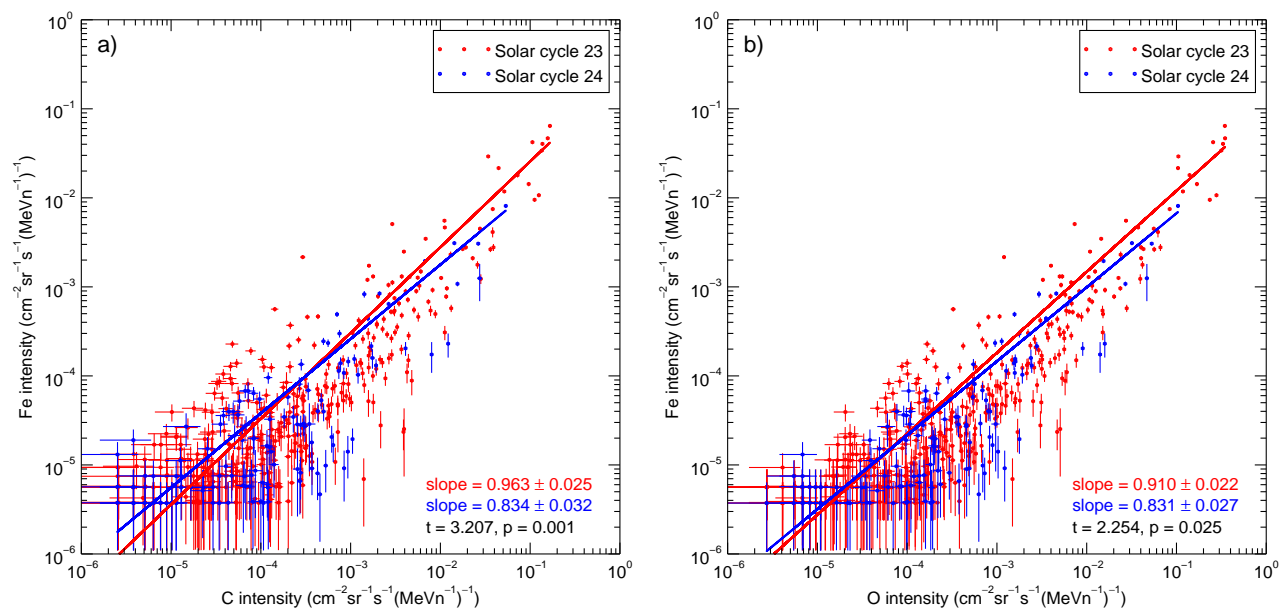}
\caption{\textbf{a)} Daily average 5--15 MeV~n$^{-1}$ Fe intensities versus daily average C intensities. \textbf{b)} Daily average 5--15 MeV~n$^{-1}$ Fe intensities versus daily average O intensities. In both figures, solar cycle 23 is shown in red and solar cycle 24 in blue. Days with fewer than two detected counts of C, O, or Fe are omitted. Linear fits in log-log scales and their slopes are shown. The quantities $t$ and $p$ represent a test statistic calculated for the difference of the slopes and the probability that such a difference would occur by chance. See Appendix \ref{sect_stattest} for details on the fitting and statistical testing.} \label{led_scatter}
\end{figure*}
We started our investigation by inspecting daily average intensities of iron, carbon, and oxygen in the energy range 5--15 MeV~n$^{-1}$. The results are presented in Sect. \ref{sect_daily_int}. We then selected Fe-rich days with a clearly higher Fe/(C+O) ratio than the average value measured for gradual solar particle events. From the selected days we identified separate Fe-rich SEP events. These results are discussed in Sect. \ref{sect_event_sel}. In Sect. \ref{sect_flares} we present properties of the solar events associated with the identified SEP events. Section \ref{sect_prop_heavy} is devoted to a detailed study of the event-integrated intensities of individual heavy elements during solar cycles 23 and 24.

\subsection{Daily intensities}\label{sect_daily_int}
Figures \ref{led_scatter}a and b show the daily average intensities of 5--15 MeV~n$^{-1}$ Fe versus the daily average intensities of 5--15 MeV~n$^{-1}$ C and O, respectively. Solar cycle 23 is shown in red and cycle 24 in blue; this coloring format is used throughout the paper. In both figures, days with fewer than two detected counts of either of the corresponding ion species are omitted. In units of intensity, the two counts correspond to about $2.6\cdot10^{-6}$~cm$^{-2}$sr$^{-1}$s$^{-1}$(MeVn$^{-1}$)$^{-1}$, $2.8 \cdot 10^{-6}$~cm$^{-2}$sr$^{-1}$s$^{-1}$(MeVn$^{-1}$)$^{-1}$ and $3.8\cdot 10^{-6}$~cm$^{-2}$sr$^{-1}$s$^{-1}$(MeVn$^{-1}$)$^{-1}$ for C, O, and Fe, respectively. The red and blue lines are linear fits in log-log scales for cycles 23 and 24. The slopes of the fits are shown, as well as a test statistic $t$ calculated from the difference of the slopes. The $p$-values in Figs. \ref{led_scatter}a and b are the results of the t-tests, representing the probabilities that the test statistic would be equal to or higher than $t$ if the samples were drawn randomly from a common distribution. The difference of slopes for cycles 23 and 24 is statistically significant within the 95\% confidence limit in both Figs. \ref{led_scatter}a and b, which means that the cycles have been significantly different considering these elements. Details on the fitting and the calculation of the test statistic are given in Appendix \ref{sect_stattest}. The overall number of days with at least two counts of both ion species in Fig. \ref{led_scatter}a is 502; 380 during solar cycle 23 and 122 during solar cycle 24. These correspond to rates of 30.3 a$^{-1}$ for cycle 23 (duration of 12.6 years) and 18.5 a$^{-1}$ for cycle 24 (observed duration of 6.6 years until the end of June 2015). In Fig. \ref{led_scatter}b the number of days is 511; 390 during cycle 23 and 121 during cycle 24, corresponding to rates of \mbox{31.1 a$^{-1}$} and \mbox{18.4 a$^{-1}$} for cycles 23 and 24, respectively.

\begin{figure*}
\centering
\includegraphics[width=17cm]{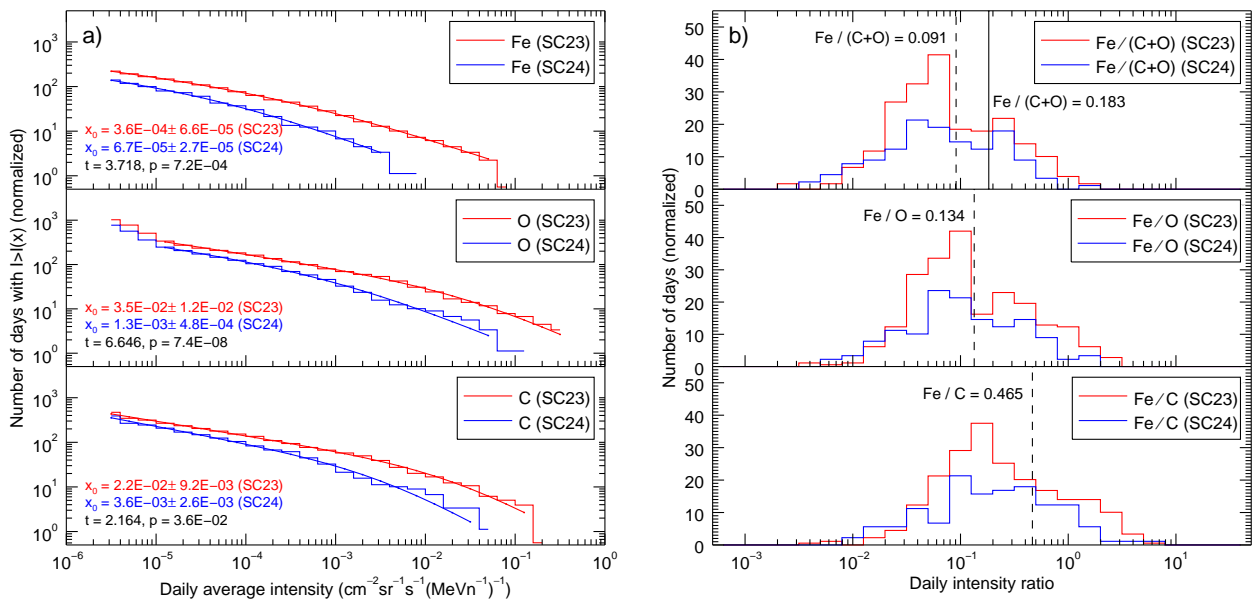}
\caption{\textbf{a)} Cumulative distribution of daily average intensities of 5--15 MeV~n$^{-1}$ C (\emph{bottom}), O (\emph{middle}) and Fe (\emph{top}). The quantities $x_0$ are the breakpoint intensities of the double power-law fits. Similarly to Fig. \ref{led_scatter}, the quantities $t$ and $p$ represent a test statistic calculated from the logarithmic difference of the breakpoint intensities and its probability value. \textbf{b)} Distribution of daily average \mbox{5--15 MeV~n$^{-1}$} intensity ratios Fe/C (\emph{bottom}), Fe/O (\emph{middle}), and Fe/(C+O) (\emph{top}). The dashed lines show the average values measured in gradual events, and the solid line in the \emph{top panel} shows the gradual event value multiplied by two. In both figures, the colors and data omission are similar as in Fig. \ref{led_scatter}. All distributions have been normalized to account for the time SOHO/ERNE has been offline, and the SC23 distributions have been scaled down to account for the difference between the duration of SC23 and the measured duration of SC24 (until 30 June 2015).} \label{led_distributions}
\end{figure*}

Figure \ref{led_distributions}a shows the cumulative distributions of daily average intensities of 5--15 MeV~n$^{-1}$ C (bottom panel), O (middle panel), and Fe (top panel) for solar cycles 23 and 24. Again, days with fewer than two counts of the corresponding ion species are omitted. This omission does not affect the shape of the distributions. All distributions were normalized to account for the amount of time SOHO/ERNE has been offline during the period in consideration. In addition, the distributions of solar cycle 23 were divided by the ratio $t_{SC23}/t_{SC24}$, where $t_{SC23}$ is the length of solar cycle 23 and $t_{SC24}$ is the length of solar cycle 24 up to June 30, 2015. The distributions were fit with double power-law functions with the breakpoint as one of the fitting parameters, and the breakpoint intensities are given in the figure. It should be noted that the data ranges used for the fitting vary between the fits; the bump in the lowest intensities of O was omitted, as was the drop in the highest intensities for all distributions except for cycle 23 O. The distributions of cycle 24 decrease more quickly than those of cycle 23 when the intensity increases, reflecting the lower solar activity, that is, the smaller number of SEP events with higher intensities. The difference of the logarithms of the power-law breakpoints were tested with t-tests similarly to Fig. \ref{led_scatter}, and the probability values from the tests are shown in the figure. The difference is significant within the 95\% confidence limit for all of the elements C, O, and Fe.

\begin{figure}
\resizebox{\hsize}{!}{\includegraphics{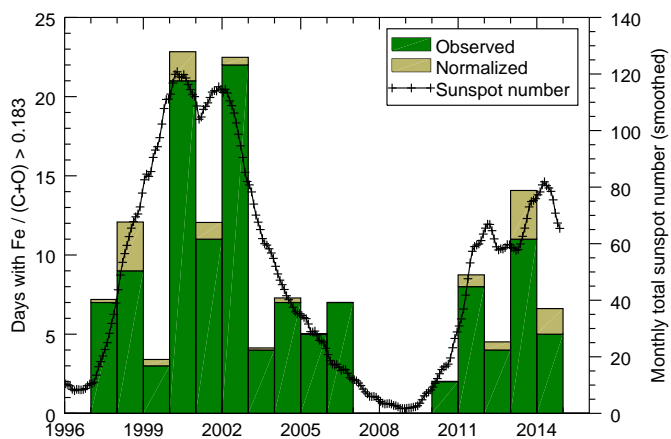}}
\caption{Annual number of days with Fe/(C+O) $>0.183$ (green bars) and the mean annual sunspot number (black curve). The normalizations (light green bars) account for the time SOHO/ERNE has been offline during each year.} \label{imp_days_per_year}
\end{figure}

Figure \ref{led_distributions}b shows the distributions of daily average 5--15 MeV~n$^{-1}$ intensity ratios Fe/C (bottom panel), Fe/O (middle panel), and Fe/(C+O) (top panel). The distributions were normalized in the same way as the intensity distributions in Fig. \ref{led_distributions}a. In each panel, a dashed line indicates the corresponding average values in gradual events measured at 5-12 MeV~n$^{-1}$ \citep{reames1995a}. The solid line in the top panel shows the value in gradual events multiplied by two; this value is used as the criterion to select Fe-rich events in Sect. \ref{sect_event_sel}. The distributions for solar cycle 24 are significantly lower than for cycle 23, mainly because of the lower number of days with two or more counts of Fe. The distributions reflect the bimodal shape reported in earlier surveys \citep[e.g.,][]{reames1988a},that is, they indicate two distinct populations of particles. The peaks of the populations with the lower values of Fe/O and Fe/(C+O) are close to the average gradual values divided by two. The corresponding peak for Fe/C is close to the gradual value divided by three. We explored the statistical validity of the observed two-peak structure by changing the binning of the data and found that it is a persistent feature of the distributions.

The daily averages in Fig. \ref{led_distributions}b suggest that cycle 23 was dominated by gradual events, whereas the number of gradual and impulsive events has been roughly equal during cycle 24. However, when we consider complete SEP events instead of just daily averages, the impulsive event population is lacking in cycle 24. This is discussed in more detail in Sect. \ref{sect_flares}. When we study daily average values like in Fig. \ref{led_distributions}b, we need to bear in mind that impulsive events have short durations, which meansthat more than one impulsive event may occur during one day. Furthermore, large gradual events have durations of several days, thus the same event may be sampled several times. Therefore the daily averages cannot be used too literally to deduce information about the number of actual SEP events, as the values attributed to gradual events are grossly overrepresented.

\begin{figure*}
\centering
\includegraphics[width=17cm]{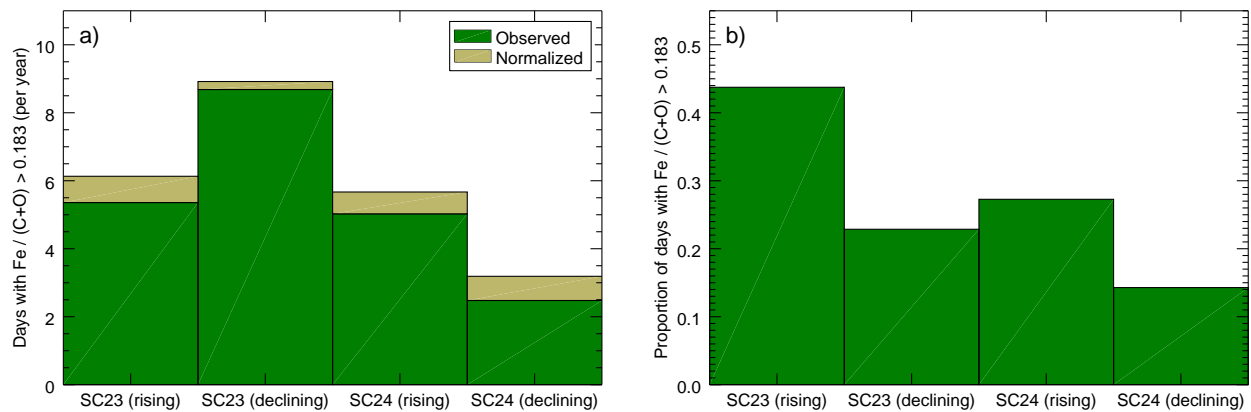}
\caption{Number (\textbf{a}) and proportion (\textbf{b}) of days with Fe/(C+O) $>0.183$ during the rising and declining phases of solar cycles 23 and 24. The normalizations in \textbf{a)} are similar as in Fig. \ref{imp_days_per_year}.} \label{imp_days_per_cycle}
\end{figure*} 

\subsection{Selection of iron-rich SEP events}\label{sect_event_sel}
We wished to compare the two solar cycles with respect to the properties of SEP events with enhanced abundances of heavy elements, therefore we used the following two criteria to search for Fe-rich events: 1) The daily counts of C, O and Fe must each be two or more, and 2) the ratio of Fe/(C+O) must be higher than $0.183$, which is twice the corresponding value of gradual solar particle events in the 5--12 MeV~n$^{-1}$ range as reported in \citet{reames1995a}. There were 126 days that fulfilled the criteria, 96 of them during solar cycle 23 and 30 during cycle 24. Figure \ref{imp_days_per_year} shows the number of days with Fe-enhancement for each year (green bars), with normalizations for the time SOHO/ERNE has been offline during each year (light green bars). The smoothed monthly international sunspot number is shown for comparison. The number of days with Fe-enhancement seems to roughly follow the mean sunspot number (an indicator of the overall solar activity), but there are also large deviations. For example, there were fewer days with Fe-enhancement during 1999, 2001, 2003, 2012, and 2014  than would have been expected from the activity level.

Figure \ref{imp_days_per_cycle}a shows the number of days with Fe-enhancement in units of a$^{-1}$ for the two cycles, divided into periods before and after the cycle maximum, with similar normalizations as in Fig. \ref{imp_days_per_year}. From this figure it is clear that the two solar cycles are different considering the heavy elements. The number of days per year with Fe-enhanced solar activity during the declining phase of the current cycle is about one third of what it was during the previous cycle at the corresponding time, while during the rising phases of the cycles the rates are roughly equal. In comparison, Fig. \ref{imp_days_per_cycle}b shows the ratio of days with Fe-enhancement (Fe/(C+O) $>0.183$) to all days with heavy ion activity (any value of Fe/(C+O) with at least two counts of each species recorded per day) during the rising and declining cycle phases. This figure shows that the portion of days with Fe-enhanced activity has been considerably smaller during the current cycle than in the previous cycle. It is interesting to note that the rising phases of both the previous and the current solar cycle had about twice as large a proportion of days with Fe-enhanced activity than the declining phases, and in this respect the cycles have been similar. The declining phase of cycle 24 is still ongoing, and the situation is thus subject to change with the detection or non-detection of Fe-enriched solar activity.

To study complete SEP events instead of regarding each day as a separate event, we made a visual scan of the SOHO/ERNE proton data and used previous catalogs of SEP events \citep{cane2010a,vainio2013a} to associate the Fe-enriched days with SEP events. In addition, we used the NOAA GOES flare database \footnote{Available online at \\ \url{ftp://ftp.ngdc.noaa.gov/STP/space-weather/solar-data/solar-features/solar-flares/x-rays/goes/}} and SOHO/LASCO CME catalog\footnote{Available online at \url{http://cdaw.gsfc.nasa.gov/CME_list/}} to obtain flare and CME associations for each event. In the end, we identified 86 events with one or more days with Fe enrichment; 65 of them occurred during solar cycle 23 and 21 during cycle 24. The event list with full details is presented in Table \ref{event_list}.

\addtocounter{table}{1}

The onset times for many of the events were taken from the SEPServer catalog \citep{vainio2013a}, where the onset times were determined with the Poisson-CUSUM-method described in \citet{huttunenheikinmaa2005a} using $55$--$80$ MeV protons. After the publication of \citet{vainio2013a}, the catalog\footnote{Available online at \url{http://server.sepserver.eu/}} has been extended to cover the time period until the end of 2014. As several small events were not seen in that energy range and thus were not included in the catalog, we used the same method but with $12.6$--$13.8$ MeV protons to calculate the onset times. The ending times of the events were defined as when the  one-hour average $12.6$--$13.8$ MeV proton intensity drops below five times the background level, or in the cases of multiple successive events, one minute before the starting time of the following event. During the observation period, the $12.6$--$13.8$ MeV proton background changed between \textasciitilde$1 \cdot 10^{-4}$~cm$^{-2}$sr$^{-1}$s$^{-1}$MeV$^{-1}$ in 2002 and \textasciitilde$2 \cdot 10^{-4}$~cm$^{-2}$sr$^{-1}$s$^{-1}$MeV$^{-1}$ in 2009. In some events the intensity rises above five times the background for only a short period of time or not at all; in these cases we used an event duration of 12 hours.

The selection criteria mean that the events on the whole are not necessarily Fe-rich, but they include at least one day with Fe-enhancement. In some large events the Fe-enhancement occurs only during the first day of the event, for example in event 37 in Table \ref{event_list} (4 November 2001), which lasted for over 12 days. This event was also mentioned by \citet{cane2003a} as an example of the type of event where the intensity-time profiles have two peaks: one close to the time of the associated flare, with relatively high Fe/O and the other during the shock passage, with low Fe/O. For some large, multi-day events, the Fe-enhancement occurs later on in the event, for example in event  52 (22 October 2003), which was associated with an eastern (N03E17) M-class flare. In this event the proton intensities rise slowly, and peak more than 24 hours after the onset, and the Fe-enhancement occurs during and after the peak intensities. In some cases it is also possible that the enhancement is caused by another smaller Fe-rich event occurring simultaneously in the background of the larger event, without causing a discernible rise in the proton intensities.

\begin{figure*}
\centering
\includegraphics[width=17cm]{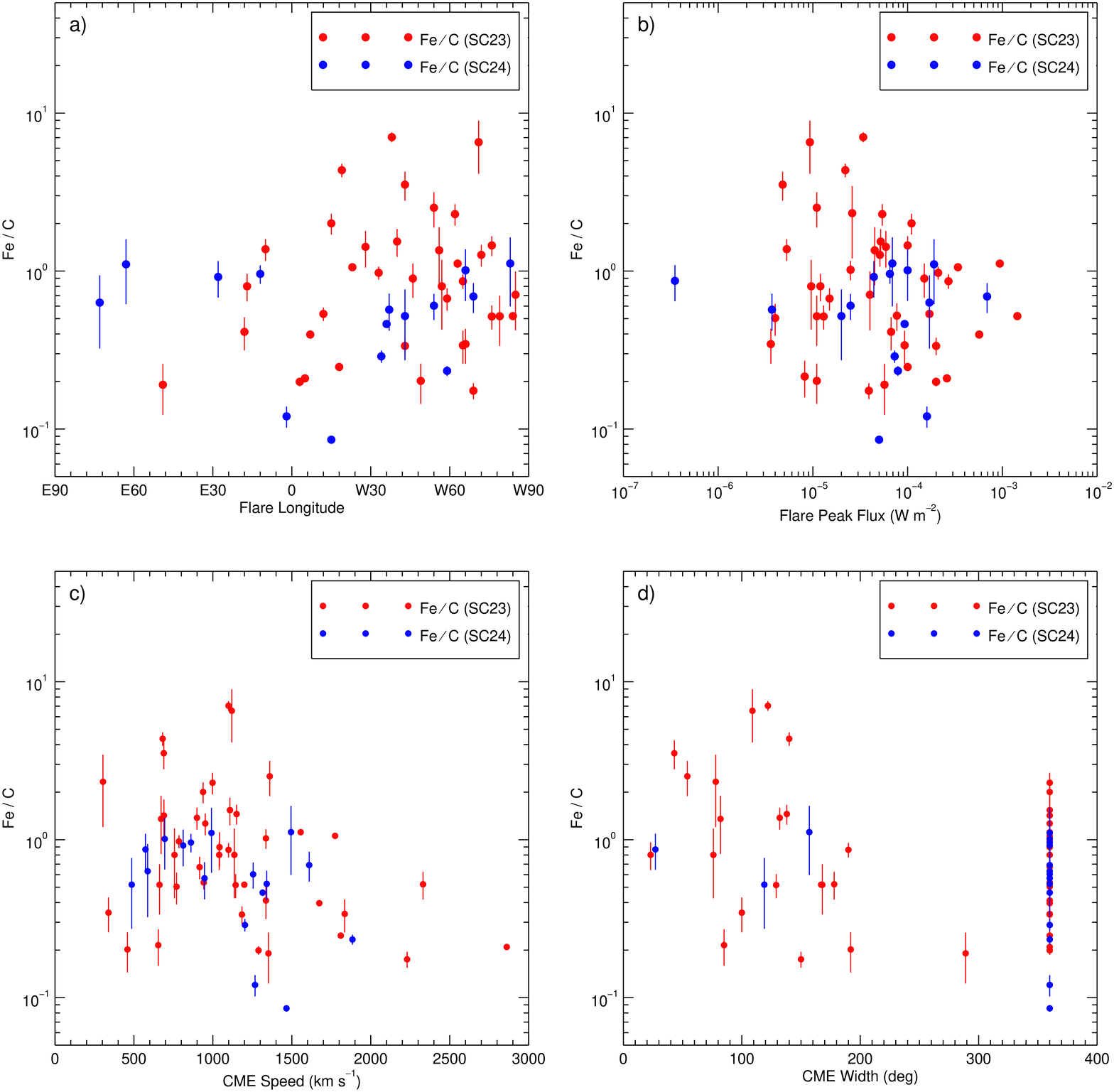}
\caption{Event-averaged 5--15 MeV~n$^{-1}$ Fe/C ratio versus longitude of the associated flare (\textbf{a}), peak X-ray intensity of the associated flare (\textbf{b}), speed of the associated CME (\textbf{c}), and width of the associated CME (\textbf{d}). In each plot, the colors are similar as in Fig. \ref{led_scatter}. Values with statistical uncertainties higher than 50\% have been omitted.} \label{fec_vs_solar_event}
\end{figure*}
\begin{figure*}
\centering
\includegraphics[width=17cm]{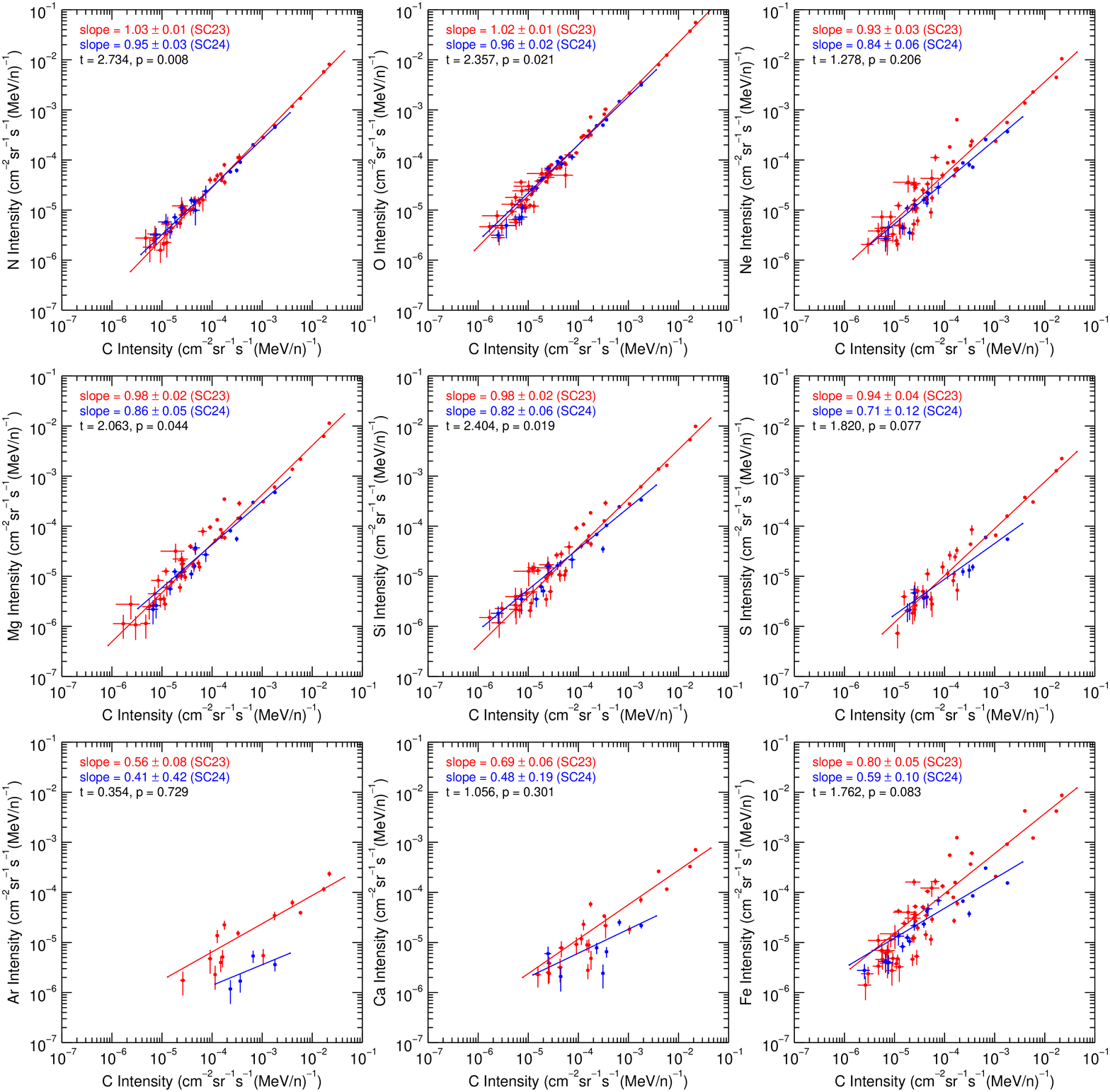}
\caption{Event-averaged 5--15 MeV~n$^{-1}$ intensities of N, O, Ne, Mg, Si, S, Ar, Ca, and Fe versus the event-averaged 5--15 MeV~n$^{-1}$ intensities of C. Each panel is shown in a similar format as in Fig. \ref{led_scatter}. Values with statistical uncertainties higher than 50\% have been omitted.}\label{x_vs_c}
\end{figure*}
\begin{figure*}
\centering
\includegraphics[width=17cm]{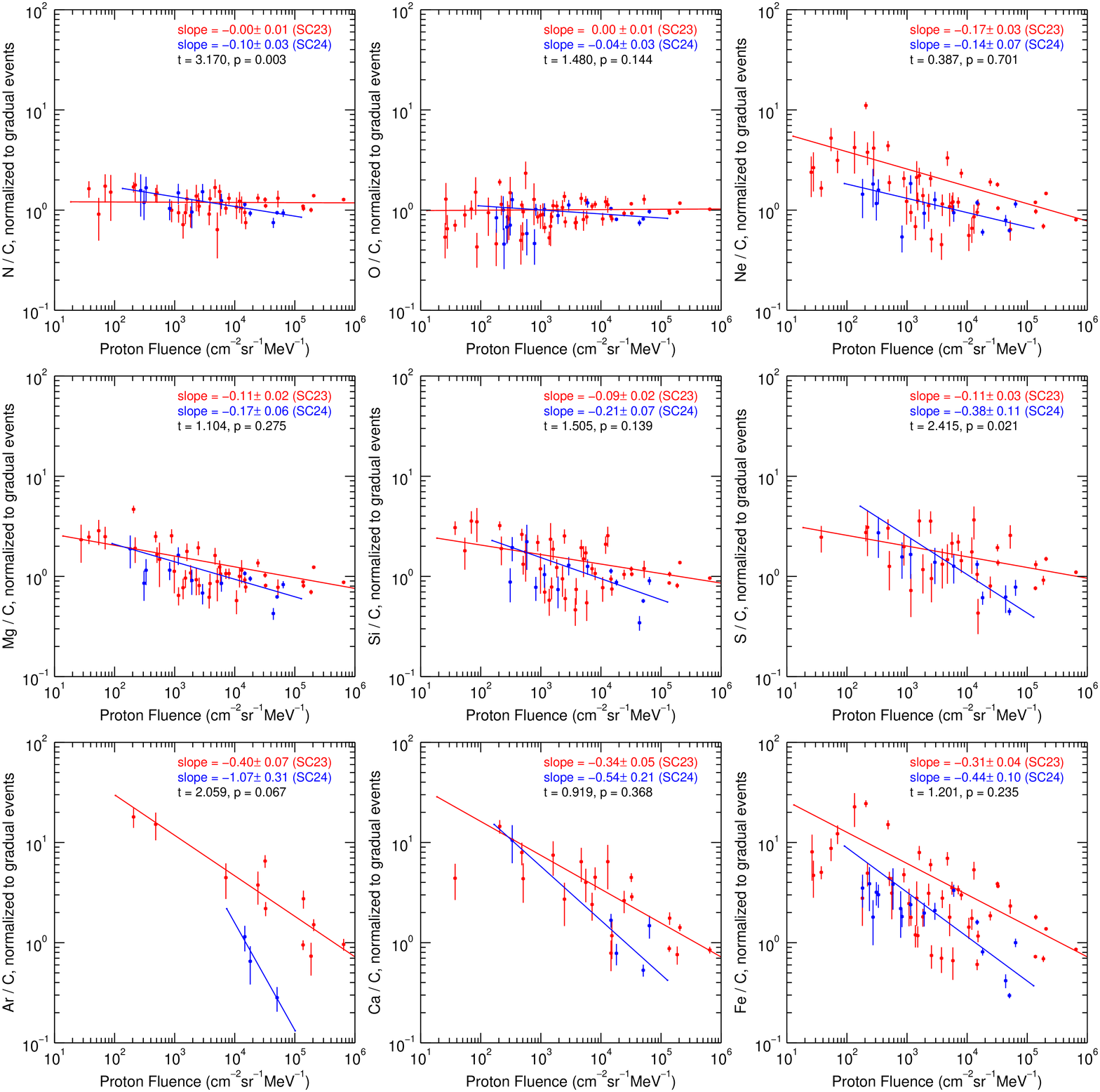}
\caption{Event-averaged ratios of 5--15 MeV~n$^{-1}$ N, O, Ne, Mg, Si, S, Ar, Ca, and Fe to C versus the event-integrated 5--15 MeV proton fluence. Each panel is shown in a similar format as in Fig. \ref{x_vs_c}.} \label{xc_vs_p}
\end{figure*}
\begin{figure*}
\centering
\includegraphics[width=17cm]{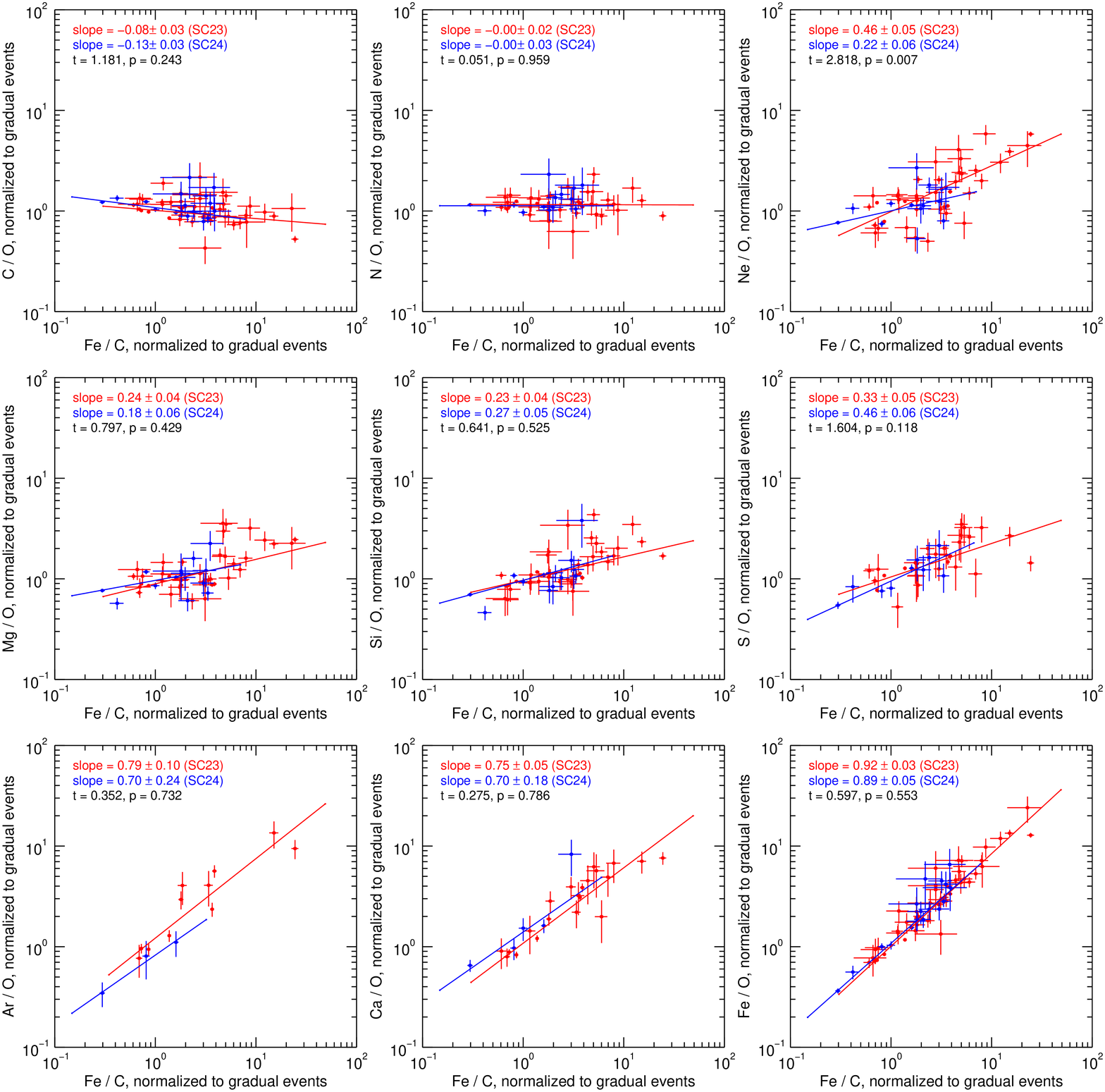}
\caption{Event-averaged ratios of 5--15 MeV~n$^{-1}$ C, N, Ne, Mg, Si, S, Ar, Ca, and Fe to O versus the event-averaged 5--15 MeV~n$^{-1}$ ratios of Fe to C. Each panel is shown in a similar format as in Fig. \ref{x_vs_c}.} \label{xo_vs_fec}
\end{figure*}

\subsection{Properties of associated flares and CMEs}\label{sect_flares}
Figure \ref{fec_vs_solar_event} investigates the possible relationships between the event-averaged 5--15 MeV~n$^{-1}$ Fe/C ratio and the properties of the associated solar flare and CME. There are different numbers of points in each panel because not all the solar event associations were found for all the events. Events with a statistical uncertainty higher than 50\% are excluded from Fig. \ref{fec_vs_solar_event} and the following analysis. In this study we chose to use Fe/C instead of the more commonly used Fe/O because C is more probably fully ionized than O; therefore Fe and C have maximally different Q/M values and their ratio is expected to exhibit the strongest effects of heavy ion enhancement.

Figure \ref{fec_vs_solar_event}a presents the Fe/C ratio versus the longitude of the associated flare. Events of solar cycle 23 with the highest Fe/C values are associated with flares at longitudes with good magnetic connection to Earth, but this behavior is not observed for events occurring during cycle 24. In fact, the two events with the highest values of Fe/C of cycle 24 (events 73 and 75) are associated with flares at N22E63 and S14W83, respectively, and especially the first is quite far from the well-connected region. The mean longitudes and their standard deviations of the associated flares are $+40 \pm 33$ for cycle 23 and $+21 \pm 48$ for cycle 24. It is also important to note that all Fe/C values for cycle 24 in the figure are below $1.2$, whereas there are many events of cycle 23 with higher Fe/C values.

In Fig. \ref{fec_vs_solar_event}b the Fe/C ratio is plotted versus the peak X-ray flux. Except for the lack of high values of Fe/C for cycle 24, there is no significant difference between the cycles. The same applies to Fig. \ref{fec_vs_solar_event}c, where the Fe/C ratio is plotted against the speed of the associated CME; both sets of events are distributed similarly. Figure \ref{fec_vs_solar_event}d shows the Fe/C ratio versus the width of the associated CME. Here the event populations are different: 82\% (14 out of 17) of the CMEs of cycle 24 in the figure are halo CMEs, compared with 50\% for cycle 23. The mean CME widths and standard deviations of the events occurring in cycles 23 and 24 are ($243 \pm 126$) degrees and ($314 \pm 104$) degrees, respectively. Recently, \citet{gopalswamy2014a,gopalswamy2015a} compared the properties of CMEs in cycles 23 and 24 and found that both the width of the non-halo CMEs and the fraction of halo CMEs have been significantly higher during solar cycle 24 than during cycle 23. They suggested that the anomalous widening of the CMEs in cycle 24 is a result of the diminished total pressure in the heliosphere.

From Figs. \ref{fec_vs_solar_event}a--d it is clear that there is a population with Fe/C > 1 among the events of cycle 23 that is almost completely lacking from the events of cycle 24. These events seem to be associated with flares with good magnetic connection and relatively low X-ray peak flux, and with relatively slow and narrow CMEs. These characteristics indicate that they are impulsive events. All but one of the events of cycle 23 in Fig. \ref{fec_vs_solar_event}a with Fe/C > 1 have rather short flare durations (shorter than an hour), giving further indication of the impulsive nature of these events, which are absent during cycle 24.

\subsection{Properties of heavy ions}\label{sect_prop_heavy}
Figure \ref{x_vs_c} shows the event-averaged 5--15 MeV~n$^{-1}$ intensities of N, O, Ne, Mg, Si, S, Ar, Ca, and Fe plotted versus the event-averaged 5--15 MeV~n$^{-1}$ intensity of C. Each panel is shown in a similar format as in Fig. \ref{led_scatter}. As in Fig. \ref{fec_vs_solar_event}, events with a statistical uncertainty higher than 50\% are excluded from this and the following figures and analyses. Linear fits on log-log scales were performed for the data. The statistical significance of the differences between the slopes of the fits for cycles 23 and 24 were tested with a t-test. At a 95\% confidence level the differences were significant for N, O, Mg, and Si, but not for other elements, even though the differences were quite large for elements heavier than Si. It should also be noted that even though the differences are not statistically significant separately, they are all in the same direction, namely the slopes for cycle 24 are smaller for each ion species. Thus, the probability is very low that all the differences, although not all significant separately at a 95\% confidence level, occur by chance.

The relation between the proton fluence and the relative abundances of heavy ions is studied in Fig. \ref{xc_vs_p}. The ratios were normalized to corresponding values in gradual events as reported in \citet{reames1995a}. Proton fluence is taken here as an indicator of the event size. Figure \ref{xc_vs_p} clearly shows that the ratio X/C decreases as the event size increases and that the decrease is stronger for heavier elements X. Even though the difference of slopes of the fits is statistically significant for only N/C and S/C, the populations are clearly different for Ne/C, Ar/C, Ca/C, and Fe/C. It is also worth noting that the events with the lowest proton fluences seem to be lacking from cycle 24 populations, again indicating the impulsive nature of the lacking events.

Figure \ref{xo_vs_fec} shows the event-averaged abundance ratios of C, N, Ne, Mg, Si, S, Ar, Ca, and Fe to O versus the ratio of Fe to C. Again, the ratios were normalized to gradual values. On average, the slopes of the fits become steeper when moving from C/O towards Fe/O. Similar results of fractionation have been reported before, for example, for X/C vs. Fe/C \citep{reames1994a} at \mbox{$1.9$--$2.8$} MeV~n$^{-1}$ and for X/O vs. Fe/C \citep{mason2004a} at 320--450 keV~n$^{-1}$. Events of cycle 23 do not exhibit this behavior as regularly as events of cycle 24. Ne/O, with a steeper-than-expected slope for both solar cycles, is an exception; it is also the only case where the difference of slopes between the solar cycles is statistically significant. In addition, Fe/C values for cycle 23 seem to be evenly distributed in log(Fe/C), as also noted by \citet{mason2004a} at lower energies. This is not true for Fe/C values for cycle 24, most of which are found at about $1.5$--4 times the gradual value.

\begin{table*}
\caption{Average heavy ion abundances for Fe-rich events.}
\label{abundances}
\centering
\begin{tabular}{ccccccc}
\hline \hline
Element & All & Cycle & Cycle & Gradual & LSEP & Corona \\
 & Events\tablefootmark{a} & 23\tablefootmark{b} & 24\tablefootmark{c} & Events\tablefootmark{d} & Events\tablefootmark{e} & ($1.4 \cdot 10^6$K)\tablefootmark{f} \\
\hline
C  & $=1.000$        & $=1.000$         & $=1.000$         & $=1.00\pm0.03$ & $=1.00\pm0.05$ & $=1.00$ \\
N  & $0.303\pm0.014$ & $0.299\pm0.017$  & $0.315\pm0.022$  & $0.27\pm0.01$  & $0.33\pm0.01$  & $0.25$  \\
O  & $2.062\pm0.080$ & $2.127\pm0.098$  & $1.864\pm0.118$  & $2.15\pm0.05$  & $2.77\pm0.11$  & $2.04$  \\
Ne & $0.541\pm0.068$ & $0.591\pm0.086$  & $0.371\pm0.037$  & $0.33\pm0.01$  & $0.42\pm0.02$  & $0.39$  \\  
Mg & $0.521\pm0.044$ & $0.552\pm0.054$  & $0.422\pm0.046$  & $0.42\pm0.01$  & $0.63\pm0.03$  & $0.46$  \\
Si & $0.422\pm0.036$ & $0.461\pm0.044$  & $0.304\pm0.052$  & $0.33\pm0.01$  & $0.65\pm0.04$  & $0.44$  \\
S  & $0.075\pm0.010$ & $0.085\pm0.012$  & $0.050\pm0.013$  & $0.07\pm0.00$  & $0.16\pm0.01$  & $0.06$  \\
Ar & $0.008\pm0.003$ & $0.010\pm0.004$  & $0.001\pm0.001$  & $0.01\pm0.00$  & \ldots         & $0.01$  \\
Ca & $0.044\pm0.010$ & $0.048\pm0.011$  & $0.029\pm0.020$  & $0.02\pm0.00$  & $0.06\pm0.01$  & $0.03$  \\
Fe & $1.104\pm0.166$ & $1.283\pm0.222$  & $0.630\pm0.080$  & $0.23\pm0.01$  & $1.12\pm0.14$  & $0.38$  \\
\hline
\end{tabular}
\tablefoot{\tablefoottext{a}{All events, this work.} 
\tablefoottext{b}{Solar cycle 23, this work.} 
\tablefoottext{c}{Solar cycle 24, this work.} 
\tablefoottext{d}{Gradual events \citep{reames1995a}.} 
\tablefoottext{e}{Large solar energetic particle (LSEP) events \citep{desai2006a}.} 
\tablefoottext{f}{Quiet corona at $1.4 \cdot 10^6$K \citep{feldman2003a}.}}
\end{table*}

Finally, Table \ref{abundances} presents the unweighted mean abundances of various ions (relative to C) in the Fe-rich events in this study for all events (Col. 2) and separately for solar cycles 23 and 24 (Cols. 3 and 4, respectively). The uncertainties are calculated as the standard error of the mean. Values for gradual events \citep{reames1995a}, large solar energetic particle (LSEP) events \citep{desai2006a}, and quiet corona \citep{feldman2003a} are given as a comparison. This table shows that the mean abundances of the heavy ions from O to Fe in the Fe-rich events of cycle 24 are significantly lower than in cycle 23. For example, the average abundance of Fe in the Fe-rich events of cycle 24 is less than half of what it was in cycle 23. Still, compared to gradual events or the coronal values, the abundance of Fe in cycle 24 was clearly enhanced. The mean abundances of all Fe-rich events in this survey were between the values of gradual events and LSEP events at least to within the error limits, except for Ne, which had a higher value than either of the compared populations.

\section{Summary and conclusions}\label{sect_concl}
We surveyed the heavy ion data measured by SOHO/ERNE from the beginning of solar cycle 23 until the end of June 2015, that is, well beyond the maximum of solar cycle 24. The long observation period allowed us to study and compare the properties of heavy ions in SEP events during the two solar cycles. We used the energy range of 5--15 MeV~n$^{-1}$ and were restricted to average daily intensities of $2.6\cdot10^{-6}$ cm$^{-2}$sr$^{-1}$s$^{-1}$(MeVn$^{-1}$)$^{-1}$, $2.8 \cdot 10^{-6}$ cm$^{-2}$sr$^{-1}$s$^{-1}$(MeVn$^{-1}$)$^{-1}$ and $3.8\cdot 10^{-6}$ cm$^{-2}$sr$^{-1}$s$^{-1}$(MeVn$^{-1}$)$^{-1}$ for carbon, oxygen, and iron, respectively. These intensities correspond to two or more counts per day recorded by the instrument in the energy range 5--15 MeV~n$^{-1}$.

The number of days per year averaged over solar cycle with C, O, and Fe intensities above the thresholds given above were significantly higher (by a factor of $\geq$ 1.6) during cycle 23 than in cycle 24. When presenting the Fe intensities as functions of C or O intensities, the populations in cycles 23 and 24 behaved differently. During cycle 23, much higher C, O, and in particular Fe daily average intensities were recorded than in cycle 24 (Fig. \ref{led_scatter}). Linear fits in log-log scales of Fe intensity as a function of C or O intensity gave steeper slopes (at 95\% confidence levels) for cycle 23 than for cycle 24. These results were the first indications of the reduced efficiency of solar particle acceleration in cycle 24.

The normalized number of days for carbon and oxygen at low cumulative daily average intensities were approximately equal during cycles 23 and 24, but days with higher intensities ($\geq$ 10$^{-3}$~cm$^{-2}$sr$^{-1}$s$^{-1}$(MeVn$^{-1}$)$^{-1}$) occurred much less frequently during cycle 24 (Fig. \ref{led_distributions}). The difference for Fe between the cycles was clear at all intensities, with a lower normalized number of days in cycle 24. When fitting the distributions with double power laws, the breakpoints for all three elements were at lower intensities during cycle 24 and the distributions in cycle 23 extended to considerably higher maximum intensities. The latter was particularly true for iron. The fewer days with high C and O intensities in cycle 24 again indicate a weaker overall acceleration efficiency of SEPs. The complete absence of days with high Fe intensities ($\geq 8 \cdot 10^{-3}$~cm$^{-2}$sr$^{-1}$s$^{-1}$(MeVn$^{-1}$)$^{-1}$) seems to imply that the processes responsible for impulsive acceleration in particular are weaker during cycle 24. The bimodal distributions of Fe/(C+O) and Fe/O ratios show, however, that there are days with both gradual and impulsive SEP events during cycle 24 as well, although fewer than in cycle 23.

The number of Fe-enhanced days, defined as Fe/(C+O) $>0.183$, was much higher during cycle 23 and there was a clear difference between the rising and declining phases of the two cycles, with the largest number of Fe-enhanced days in the declining phase of cycle 23 (Fig. \ref{imp_days_per_cycle}a). When calculating the portion of Fe-enhanced days from all days with C, O, and Fe intensities above the thresholds (i.e., with any value of the Fe/(C+O) ratio and with at least two counts of each species recorded per day), both cycles had about twice as large portions of Fe-enhanced days in the rising phase than in the declining phase. Thus, although during the declining phase of cycle 23 there was a significantly larger number of days with C, O, and Fe detected than the rising phase, relatively few of these were Fe-enhanced. From comparing the cycles, we found the portions of Fe-enhanced days in both rising and declining phases of cycle 23 to be about twice larger than during cycle 24 at corresponding times (Fig. \ref{imp_days_per_cycle}b). This may be due to the significantly lower number of M- and X-class flares in cycle 24.

When identifying complete SEP events instead of individual days, 86 SEP events with Fe enrichment were found (Table \ref{event_list}). Of these, 65 occurred in cycle 23 and 21 in cycle 24. For the identified Fe-rich SEP events, we investigated the dependence of the Fe/C ratios on the flare longitude, X-ray flare peak flux, CME speed, and CME width (Fig. \ref{fec_vs_solar_event}). While the events of cycle 23 with highest Fe/C ratios were associated with flares at western longitudes, this was not observed during cycle 24, which had a much flatter flare longitude distribution. In this respect, cycle 23 events obeyed the expected behavior of Fe-rich impulsive SEP events originating from magnetically well-connected regions. In cycle 24, however, particles of Fe-rich events seem to have had easier access from distant longitudes to field lines connecting to Earth.  No differences were observed between the cycles in the Fe/C ratio dependence on X-ray flare peak flux or CME speed. With respect to the CME widths, the SEP populations of the two cycles were different. In cycle 24, 82\% of the Fe-rich SEP events for which a CME association was found were associated with halo CMEs. The corresponding portion in cycle 23 was only 50\%. This may be related to the generally more frequent appearance of halo CMEs during cycle 24, presumably because of the diminished total pressure in the heliosphere \citep{gopalswamy2015a}. Associations of eastern Fe-rich events with rapidly expanding CMEs may also explain the access of particles to field lines connecting to Earth.

The population of events with Fe/C $>1$, which was present during cycle 23, was completely lacking during cycle 24 (Fig. \ref{fec_vs_solar_event}). The characteristics of these events during cycle 23 implied that they were impulsive events. The absence of these events in cycle 24 may be due to a reduced particle acceleration efficiency or to a different or less abundant seed particle populations. The latter may be related to a decrease in the efficiency of plasma fractionation processes due to low level of solar activity \citep{landi2015a}.

The behavior of intensities of N, O, Mg, and Si vs. the intensity of C was found to be different during cycles 23 and 24. With 95\% confidence levels, the slopes of the log-log fits for these elements were steeper during cycle 23 (Fig. \ref{x_vs_c}). This is caused by the presence of large SEP events with high intensities of various elements in cycle 23, which were not present in cycle 24. This indicates a lower efficiency in the shock acceleration in cycle 24 that might be due to reduced turbulence in the low corona and in interplanetary space.

With increasing proton fluences of the events, the event-averaged X/C ratios were decreasing, and the decrease was stronger for heavier elements (Fig. \ref{xc_vs_p}). No systematic statistically significant difference between the cycles was found, therefore we cannot reliably estimate whether there has been a change between the cycles in the efficiency of an A/Q-dependent acceleration process, for instance. For several elements, however, the populations during the two cycles were clearly different, and the events with the lowest proton fluences were lacking from the cycle 24 populations.

The event-averaged abundance ratios of X/O vs. Fe/C were found to generally increase and the slopes of the linear fits in log-log scales became steeper when moving from C/O toward Fe/O (Fig. \ref{xo_vs_fec}). This was true for both cycles, and there was no statistically significant difference in the slopes between the cycles, excluding Ne/O, which had a steeper slope during cycle 23, and had a steeper slope than expected from the systematics of the other elements during cycle 24 as well. The values of Fe/C during cycle 23 were evenly distributed on a logarithmic scale, while during cycle 24 the ratio distribution was more restricted closer to the coronal value.

The mean abundances of heavy ions from O to Fe in Fe-rich SEP events were found to be significantly lower for cycle 24 than for cycle 23 (Table \ref{abundances}). Compared to gradual SEP events or to coronal values, however, the abundance of Fe in these events was found to be clearly enhanced even during cycle 24.

Overall, we found that the properties of heavy ions in solar energetic particle events reflect the reduced solar activity and possibly the weaker magnetic field strength during solar cycle 24. There were fewer solar particle events with Fe intensities exceeding our threshold value in the energy range 5--15 MeV~n$^{-1}$ during solar cycle 24 than during cycle 23. Furthermore, in general the average heavy ion abundances of cycle 24 events were lower than during cycle 23. Our results indicate lower efficiencies of solar particle acceleration processes in both large SEP events and weaker impulsive events during cycle 24 and possibly differences in the composition and abundance of seed particles between the two cycles.

\begin{acknowledgements}
We gratefully acknowledge the use of data made available at the GSFC and NOAA STP online services. The CME catalog is generated and maintained at the CDAW Data Center by NASA and The Catholic University of America in cooperation with the Naval Research Laboratory. SOHO is a project of international cooperation between ESA and NASA. O. R. wishes to thank the Vilho, Yrj\"o and Kalle V\"ais\"al\"a foundation for financial support. We thank the referee, Eberhard M\"obius, for valuable comments, which improved the manuscript significantly.
\end{acknowledgements}

\bibliographystyle{aa}
\bibliography{oajraubib}
\clearpage
\onecolumn
\begin{longtab} \small
\setcounter{table}{1}
\begin{longtable}{cccc ccccc cccc}
\caption{List of SEP events with Fe-enhancements.}\label{event_list} \\
\hline \hline
& \multicolumn{3}{c}{SEP\tablefootmark{a}} & \multicolumn{5}{c}{Flare\tablefootmark{b}} & \multicolumn{3}{c}{CME\tablefootmark{c}} &  \\
\cmidrule(r){2-4} \cmidrule(lr){5-9} \cmidrule(l){10-12}
ID & Date & Time & Dur. & Start & Max. & End & Pos. & Class & Start & Width & Speed & Fe/C\tablefootmark{d} \\
& & & (h) & & & & & & & (deg) & (kms$^{-1}$) & \\
\hline
\endfirsthead
\caption{Continued.} \\
\hline \hline
& \multicolumn{3}{c}{SEP\tablefootmark{a}}  & \multicolumn{5}{c}{Flare\tablefootmark{b}} & \multicolumn{3}{c}{CME\tablefootmark{c}} &  \\
\cmidrule(r){2-4} \cmidrule(lr){5-9} \cmidrule(l){10-12}
ID & Date & Time & Dur. & Start & Max. & End & Pos. & Class & Start & Width & Speed & Fe/C\tablefootmark{d} \\
& & & (h) & & & & & & & (deg) & (kms$^{-1}$) & \\
\hline
\endhead
\hline
\endfoot
1  & 1997--Nov--4  & 06:41 & 54  & 05:52  & 05:58  & 06:02  & S14W33  & X2.1   & 06:10  & 360    & 785    & $0.98\pm0.09$ \\
2  & 1997--Nov--6  & 12:37 & 178 & 11:49  & 11:55  & 12:01  & S18W63  & X9.4   & 12:10  & 360    & 1556   & $1.12\pm0.04$ \\
3  & 1998--May--2  & 14:10 & 90  & 13:31  & 13:42  & 13:51  & S15W15  & X1.1   & 14:06  & 360    & 938    & $2.00\pm0.30$ \\
4  & 1998--May--6  & 08:29 & 68  & 07:58  & 08:09  & 08:20  & S11W65  & X2.7   & 08:29  & 190    & 1099   & $0.86\pm0.09$ \\
5  & 1998--May--9  & 04:32 & 128 & 03:04  & 03:40  & 03:55  & S,W100 & M7.7   & 03:35  & 178    & 2331   & $0.52\pm0.10$ \\
6  & 1998--May--27 & 14:48\tablefootmark{e} & 47  & 13:30\tablefootmark{g}  & 13:35  & 14:50  & N18W58  & C7.5   & 13:45  & 268    & 878    & $0.53\pm0.31$ \\
7  & 1998--Oct--18 & 22:22 & 55  & \ldots & \ldots & \ldots & N,W120 & \ldots & dg     & dg     & dg     & $0.90\pm0.40$ \\
8  & 1998--Nov--14 & 06:16 & 10  & \ldots & \ldots & \ldots & N,W120 & \ldots & dg     & dg     & dg     & $1.73\pm0.24$ \\
9  & 1999--May--27 & 11:16 & 129 & 11:36\tablefootmark{g}  & 11:43  & 11:54  & S30E78  & C4.5   & 11:06\tablefootmark{h}  & 360    & 1691   & $0.42\pm0.34$ \\
10 & 1999--Jun--4  & 15:46\tablefootmark{e} & 153 & 06:52  & 07:03  & 07:11  & N17W69  & M3.9   & 07:27  & 150    & 2230   & $0.17\pm0.02$ \\
11 & 1999--Dec--28 & 02:58\tablefootmark{e} & 51  & 00:39  & 00:48  & 00:52  & N20W56  & M4.5   & 00:54  & 82     & 672    & $1.35\pm0.54$ \\
12 & 2000--Feb--18 & 09:57 & 103 & \ldots & \ldots & \ldots & N,W120 & \ldots & 09:54  & 118   & 890    & $3.70\pm3.10$ \\
13 & 2000--Mar--8  & 01:31\tablefootmark{e} & 12  & 16:01\tablefootmark{g}  & 16:07  & 16:13  & S22E77  & M1.2   & 16:30\tablefootmark{h}  & 108    & 644    & $1.46\pm1.03$ \\
14 & 2000--May--1  & 11:29\tablefootmark{e} & 12  & 10:16  & 10:27  & 10:34  & N20W54  & M1.1   & 10:54  & 54     & 1360   & $2.52\pm0.63$ \\
15 & 2000--May--4  & 12:40\tablefootmark{e} & 33  & 10:57  & 11:08  & 11:14  & S20W90  & M6.8   & 11:26  & 170    & 1404   & $1.97\pm1.50$ \\
16 & 2000--Jun--4  & 13:06\tablefootmark{e} & 47  & 06:24\tablefootmark{g}  & 06:30  & 06:34  & N21E45  & C3.1   & 07:31\tablefootmark{h}  & 17     & 597    & $1.47\pm1.47$ \\
17 & 2000--Jun--10 & 17:26 & 124 & 16:40  & 17:02  & 17:19  & N22W40  & M5.2   & 17:08  & 360    & 1108   & $1.54\pm0.31$ \\
18 & 2000--Jun--15 & 21:00\tablefootmark{e} & 33  & 19:38  & 19:57  & 20:19  & N20W62  & M1.8   & 20:06\tablefootmark{h}  & 116    & 1081   & $0.92\pm0.53$ \\
19 & 2000--Jun--18 & 02:29 & 56  & 01:52  & 01:59  & 02:03  & N23W85  & X1.0   & 02:10  & 132    & 629    & $0.26\pm0.14$ \\
20 & 2000--Jun--23 & 15:06\tablefootmark{e} & 39  & 14:18  & 14:31  & 14:46  & N23W72  & M3.0   & 14:54  & 198    & 847    & $1.25\pm0.70$ \\
21 & 2000--Jul--11 & 01:00\tablefootmark{e} & 82  & 21:05\tablefootmark{g}  & 21:42  & 22:27  & N18E49  & M5.7   & 21:50\tablefootmark{h}  & 289    & 1352   & $0.19\pm0.07$ \\
22 & 2000--Jul--14 & 10:37 &     & 10:03  & 10:24  & 10:43  & N22W07  & X5.7   & 10:54  & 360    & 1674   & $0.40\pm0.01$ \\
23 & 2000--Aug--12 & 11:25\tablefootmark{e} & 106 & 09:45  & 09:56  & 10:09  & S17W79  & M1.1   & 10:35  & 168    & 662    & $0.52\pm0.18$ \\
24 & 2000--Sep--19 & 12:08\tablefootmark{e} & 128 & 08:06  & 08:26  & 08:42  & N14W46  & M5.1   & 08:50  & 76     & 766    & $0.54\pm0.31$ \\
25 & 2000--Oct--16 & 07:39 & 150 & 06:40  & 07:28  & 09:11  & N,W95  & M2.5   & 07:27  & 360    & 1336   & $1.02\pm0.14$ \\
26 & 2000--Oct--25 & 12:40 & 102 & 08:45  & 11:25  & 15:21  & S,W120 & C4.0   & 08:26  & 360    & 770    & $0.50\pm0.11$ \\
27 & 2000--Oct--30 & 12:47\tablefootmark{e} & 12  & \ldots\tablefootmark{g} & \ldots & \ldots & \ldots  & \ldots & dg\tablefootmark{h}     & dg     & dg     & $2.21\pm2.02$ \\
28 & 2000--Nov--24 & 05:43 & 115 & 04:55  & 05:02  & 05:08  & N22W03  & X2.0   & 05:30  & 360    & 1289   & $0.20\pm0.01$ \\
29 & 2001--Jan--28 & 16:58 & 199 & 15:40  & 16:00  & 16:24  & S04W59  & M1.5   & 15:54  & 360    & 916    & $0.67\pm0.11$ \\
30 & 2001--Mar--10 & 07:39\tablefootmark{e} & 72  & 04:00  & 04:05  & 04:07  & N27W42  & M6.7   & 04:26  & 81     & 819    & $0.34\pm0.22$ \\
31 & 2001--Mar--29 & 11:49 & 97  & 09:57  & 10:15  & 10:32  & N16W12  & X1.7   & 10:26  & 360    & 942    & $0.54\pm0.05$ \\
32 & 2001--Apr--2  & 12:24 & 10  & 10:58  & 11:36  & 12:05  & N16W62  & X1.1   & 11:26  & 80     & 992    & $1.24\pm0.64$ \\
33 & 2001--Apr--12 & 11:01 & 63  & 09:36\tablefootmark{g}  & 10:28  & 10:49  & S19W43  & X2.0   & 10:31  & 360    & 1184   & $0.34\pm0.04$ \\
34 & 2001--Apr--15 & 14:05 & 61  & 13:19  & 13:50  & 13:55  & S20W84  & X14.4  & 14:06  & 167    & 1199   & $0.52\pm0.02$ \\
35 & 2001--Sep--11 & 04:14\tablefootmark{e} & 45  & 00:49\tablefootmark{g}  & 01:11  & 01:23  & \ldots  & M2.6   & 01:55\tablefootmark{h}  & 78     & 304    & $2.33\pm1.12$ \\
36 & 2001--Oct--22 & 15:51 & 227 & 14:27  & 15:08  & 15:31  & S21E18  & M6.7   & 15:06  & 360    & 1336   & $0.41\pm0.10$ \\
37 & 2001--Nov--4  & 16:45 & 304 & 16:03  & 16:20  & 16:57  & N06W18  & X1.0   & 16:35  & 360    & 1810   & $0.25\pm0.01$ \\
38 & 2002--Jan--27 & 13:38 & 134 & \ldots & \ldots & \ldots & N,W120 & \ldots & 12:30  & 360    & 1136   & $0.80\pm0.38$ \\
39 & 2002--Feb--20 & 05:58 & 109 & 05:52  & 06:12  & 06:16  & N12W72  & M5.1   & 06:30  & 360    & 952    & $1.27\pm0.20$ \\
40 & 2002--Apr--14 & 10:15\tablefootmark{e} & 69  & 07:28  & 07:39  & 07:44  & N19W57  & C9.6   & 07:50  & 76     & 757    & $0.80\pm0.38$ \\
41 & 2002--May--30 & 06:39\tablefootmark{e} & 51  & 04:24  & 05:32  & 06:13  & N,W100 & M1.3   & 05:06  & 144    & 1625   & $1.10\pm0.84$ \\
42 & 2002--Jul--19 & 07:17\tablefootmark{e} & 70  & 23:08\tablefootmark{g}  & 23:17  & 23:23  & \ldots  & C8.2   & 01:32\tablefootmark{h}  & 85     & 654    & $0.21\pm0.06$ \\ 
43 & 2002--Aug--3  & 23:13\tablefootmark{e} & 33  & 18:59  & 19:07  & 19:11  & S16W76  & X1.0   & 19:32  & 138    & 1150   & $1.45\pm0.21$ \\
44 & 2002--Aug--5  & 07:58\tablefootmark{e} & 85  & 04:21\tablefootmark{g}  & 05:17  & 05:33  & S10W43  & C4.8   & 07:32  & 43     & 689    & $3.53\pm0.73$ \\
45 & 2002--Aug--18 & 22:10 & 35  & 21:12  & 21:25  & 21:37  & S12W19  & M2.2   & 21:54  & 140    & 682    & $4.35\pm0.42$ \\
46 & 2002--Aug--20 & 08:46 & 42  & 08:22  & 08:26  & 08:30  & S10W38  & M3.4   & 08:54  & 122    & 1099   & $7.05\pm0.50$ \\
47 & 2002--Aug--22 & 02:30 & 47  & 01:47  & 01:57  & 02:05  & S07W62  & M5.4   & 02:06  & 360    & 998    & $2.29\pm0.36$ \\
48 & 2002--Oct--30 & 04:03\tablefootmark{e} & 240 & 02:53\tablefootmark{g}  & 02:58  & 03:11  & N30W66  & C3.6   & 05:50\tablefootmark{h}  & 100    & 339    & $0.34\pm0.09$ \\
49 & 2002--Nov--26 & 19:09\tablefootmark{e} & 92  & 18:26\tablefootmark{g}  & 18:35  & 18:39  & N26W87  & C3.6   & \ldots\tablefootmark{h} & \ldots & \ldots & $0.87\pm0.80$ \\
50 & 2003--May--31 & 02:56 & 74  & 02:13  & 02:24  & 02:40  & S07W65  & M9.3   & 02:30  & 360    & 1835   & $0.34\pm0.09$ \\
51 & 2003--Aug--19 & 09:02\tablefootmark{e} & 36  & 07:38  & 07:59  & 08:01  & S12W64  & M2.0   & 08:30  & 35     & 412    & $0.29\pm0.18$ \\
52 & 2003--Oct--22 & 17:40\tablefootmark{e} & 96  & 15:57\tablefootmark{g}  & 16:01  & 16:04  & N03E17  & M1.2   & 16:30\tablefootmark{h}  & 23     & 1040   & $0.80\pm0.16$ \\
53 & 2004--Jul--22 & 17:33 & 71  & 07:41  & 07:59  & 08:08  & N04E10  & C5.3   & 08:30  & 132    & 899    & $1.38\pm0.22$ \\
54 & 2004--Oct--30 & 07:58\tablefootmark{e} & 7   & 06:08\tablefootmark{g}  & 06:18  & 06:22  & N13W22  & M4.2   & 06:54  & 360    & 422    & $2.14\pm1.38$ \\
55 & 2004--Oct--30 & 14:43\tablefootmark{e} & 3   & 11:38\tablefootmark{g}  & 11:46  & 11:50  & N13W25  & X1.2   & 12:30  & 360    & 427    & $2.19\pm1.16$ \\
56 & 2004--Oct--30 & 18:01\tablefootmark{e} & 36  & 16:18\tablefootmark{g}  & 16:33  & 16:37  & N13W28  & M5.9   & 16:54  & 360    & 690    & $1.42\pm0.37$ \\
57 & 2004--Nov--1  & 06:15 & 155 & 03:04\tablefootmark{g}  & 03:22  & 03:26  & N12W49  & M1.1   & 03:54\tablefootmark{h}  & 192    & 459    & $0.20\pm0.06$ \\
58 & 2005--Jan--15 & 23:35 & 289 & 22:25  & 23:02  & 23:31  & N15W05  & X2.6   & 23:07\tablefootmark{h}  & 360    & 2861   & $0.21\pm0.00$ \\
59 & 2005--May--6  & 02:57\tablefootmark{e} & 11  & 03:05  & 03:14  & 03:21  & S04W71  & C9.3   & 03:30  & 109    & 1120   & $6.54\pm2.41$ \\
60 & 2005--May--6  & 14:06\tablefootmark{e} & 70  & 11:11  & 11:28  & 11:35  & S04W76  & M1.3   & 11:54  & 129    & 1144   & $0.52\pm0.09$ \\
61 & 2005--Jun--16 & 20:35 & 134 & 20:01  & 20:22  & 20:42  & N09W85  & M4.0   & dg     & dg     & dg     & $0.71\pm0.29$ \\
62 & 2005--Aug--29 & 14:28 & 65  & \ldots & \ldots & \ldots & S,W120 & \ldots & 10:54  & 360    & 1600   & $0.31\pm0.17$ \\
63 & 2006--Nov--21 & 20:36\tablefootmark{e} & 12  & \ldots & \ldots & \ldots & S,W120 & \ldots & dg     & dg     & dg     & $2.95\pm2.55$ \\
64 & 2006--Dec--13 & 02:59 & 35  & 02:14  & 02:40  & 02:57  & S06W23  & X3.4   & 02:54  & 360    & 1774   & $1.06\pm0.03$ \\
65 & 2006--Dec--14 & 23:08 & 206 & 21:07  & 22:15  & 22:26  & S07W46  & X1.5   & 22:30  & 360    & 1042   & $0.90\pm0.22$ \\
66 & 2010--Jun--12 & 02:43 & 68  & 00:30\tablefootmark{g}  & 00:57  & 01:02  & N23W43  & M2.0   & 01:32\tablefootmark{h}  & 119    & 486    & $0.52\pm0.25$ \\
67 & 2010--Sep--1  & 01:28 & 23  & 21:50\tablefootmark{g}  & 21:53  & 21:56  & \ldots  & B1.8   & 21:17\tablefootmark{h}  & 360    & 1304   & $5.19\pm4.16$ \\
68 & 2011--Mar--21 & 03:27 & 137 & \ldots\tablefootmark{g} & \ldots & \ldots & \ldots  & \ldots & 02:24\tablefootmark{h}  & 360    & 1341   & $0.53\pm0.11$ \\
69 & 2011--Jun--5  & 05:10 & 50  & 02:11\tablefootmark{g}  & 02:14  & 02:17  & \ldots  & B3.5   & 03:00\tablefootmark{h}  & 27     & 573    & $0.87\pm0.22$ \\
70 & 2011--Jun--7  & 07:36 & 80  & 06:16\tablefootmark{g}  & 06:41  & 06:59  & S21W54  & M2.5   & 06:49\tablefootmark{h}  & 360    & 1255   & $0.60\pm0.11$ \\
71 & 2011--Aug--4  & 04:40 & 110 & 03:41\tablefootmark{g}  & 03:57  & 04:04  & N19W36  & M9.3   & 04:12\tablefootmark{h}  & 360    & 1315   & $0.46\pm0.02$ \\
72 & 2011--Aug--9  & 08:22 & 122 & 07:48\tablefootmark{g}  & 08:05  & 08:08  & N17W69  & X6.9   & 08:12\tablefootmark{h}  & 360    & 1610   & $0.69\pm0.15$ \\
73 & 2011--Nov--3  & 23:39 & 155 & 20:16\tablefootmark{g}  & 20:27  & 20:32  & N22E63  & X1.9   & 23:30\tablefootmark{h}  & 360    & 991    & $1.10\pm0.49$ \\
74 & 2012--Mar--13 & 17:53 & 173 & 17:12\tablefootmark{g}  & 17:41  & 18:25  & N19W59  & M7.9   & 17:36\tablefootmark{h}  & 360    & 1884   & $0.23\pm0.02$ \\
75 & 2012--Jul--8  & 23:59\tablefootmark{f} & 89  & 16:23\tablefootmark{g}  & 16:32  & 16:42  & S14W83  & M6.9   & 16:54\tablefootmark{h}  & 157    & 1495   & $1.12\pm0.52$ \\
76 & 2012--Sep--28 & 06:31\tablefootmark{f} & 129 & 23:36\tablefootmark{g}  & 23:57  & 00:34  & N06W37  & C3.7   & 00:12\tablefootmark{h}  & 360    & 947    & $0.57\pm0.15$ \\
77 & 2013--Apr--11 & 08:10 & 106 & 06:55\tablefootmark{g}  & 07:16  & 07:29  & N09E12  & M6.5   & 07:24\tablefootmark{h}  & 360    & 861    & $0.96\pm0.13$ \\
78 & 2013--May--13 & 12:56\tablefootmark{e} & 40  & 01:53\tablefootmark{g}  & 02:17  & 02:32  & N11E89  & X1.7   & 02:00\tablefootmark{h}  & 360    & 1270   & $1.48\pm1.21$ \\
79 & 2013--May--22 & 13:47 & 227 & 13:08\tablefootmark{g}  & 13:32  & 14:08  & S18W15  & M5.0   & 13:26\tablefootmark{h}  & 360    & 1466   & $0.09\pm0.00$ \\
80 & 2013--Jun--28 & 05:48\tablefootmark{e} & 46  & 01:36\tablefootmark{g}  & 01:59  & 02:28  & S16E14  & C4.4   & 02:00\tablefootmark{h}  & 360    & 1037   & $0.27\pm0.21$ \\
81 & 2013--Oct--25 & 13:49 & 64  & 07:53\tablefootmark{g}  & 08:01  & 08:09  & S08E73  & X1.7   & 08:12\tablefootmark{h}  & 360    & 587    & $0.63\pm0.31$ \\
82 & 2013--Oct--28 & 06:19 & 12  & 01:41\tablefootmark{g}  & 02:03  & 02:12  & N04W66  & X1.0   & 02:24\tablefootmark{h}  & 360    & 695    & $1.01\pm0.36$ \\
83 & 2013--Oct--28 & 18:49 & 16  & 15:07\tablefootmark{g}  & 15:15  & 15:21  & S08E28  & M4.4   & 15:36\tablefootmark{h}  & 360    & 812    & $0.92\pm0.24$ \\
84 & 2014--Apr--18 & 13:42 & 157 & 12:31\tablefootmark{g}  & 13:03  & 13:20  & S20W34  & M7.3   & 13:26\tablefootmark{h}  & 360    & 1203   & $0.29\pm0.03$ \\
85 & 2014--May--7  & 19:15\tablefootmark{e} & 33  & 16:07\tablefootmark{g}  & 16:29  & 17:03  & N15E50  & M1.2   & 16:24\tablefootmark{h}  & 360    & 923    & $0.74\pm0.64$ \\
86 & 2014--Sep--10 & 19:28 & 104 & 17:21\tablefootmark{g}  & 17:45  & 18:20  & N14E02  & X1.6   & 18:00\tablefootmark{h}  & 360    & 1267   & $0.12\pm0.02$ \\
\end{longtable}
\tablefoot{\tablefoottext{a}{Date and time of the proton event onset from the SEPServer catalog \citep{vainio2013a}, unless otherwise indicated.} \\ 
\tablefoottext{b}{X-ray flare identification from \citet{cane2010a} with additional information from NOAA GOES X-ray flare database, unless otherwise indicated.} \\
\tablefoottext{c}{CME information from \citet{cane2010a} unless otherwise indicated, except for the width, which for all events is adopted from the SOHO/LASCO CME catalog. A gap in the LASCO observations is marked by ''dg''.} \\
\tablefoottext{d}{Event-averaged Fe/C ratio.} \\
\tablefoottext{e}{Time of the proton onset determined with the Poisson-CUSUM-method described in \citet{huttunenheikinmaa2005a}, using $12.6$--$13.8$ MeV protons.} \\
\tablefoottext{f}{Proton event onset during a SOHO/ERNE data gap; onset time determined as the first minute after the gap.} \\
\tablefoottext{g}{X-ray flare identified based on information from the NOAA GOES X-ray flare database.} \\
\tablefoottext{h}{CME identified based on information from the SOHO/LASCO CME catalog.}}
\end{longtab}
\normalsize

\twocolumn
\begin{appendix}
\section{Statistical testing of the linear fits}\label{sect_stattest}
The linear fits in this paper were calculated with the procedure \texttt{fitexy} for Interactive Data Language (IDL). The procedure is a part of the widely used IDL Astronomy User's Library\footnote{Available online at http://idlastro.gsfc.nasa.gov/}. The procedure calculates a linear least-squares approximation taking into account errors in both variables, $\sigma_x$ and $\sigma_y$, by minimizing the quantity
\begin{equation}\label{chi}
\chi^2 = \displaystyle\sum_{i=0}^{N-1}\frac{(y_i-a-bx_i)^2}{\sigma_{yi}^2+b^2\sigma_{xi}^2},
\end{equation}
where $a$ is the intercept and $b$ the slope of the resulting fit. In addition to $\chi^2$, $a,$ and $b$, the procedure calculates the error estimates $\sigma_a$ and $\sigma_b$ for the fit parameters.

We wish to test whether the difference of two slopes of linear fits is statistically significant. This can be achieved using the Student t-test with the statistic
\begin{equation}
t = \frac{b_1 - b_2}{\sqrt{\sigma_{b_1}^2 + \sigma_{b_2}^2}},
\end{equation}
where $b_1$ and $b_2$ are the slopes of the two fits and $\sigma_{b_1}$ and $\sigma_{b_2}$ their errors. When the scatter of the data is large compared to the error limits of the data points, the fit is ''poor'', that is, $\chi^2$ is large, although a linear model can still be the correct model. In this case, the error estimates of the fit parameters are not meaningful. To achieve a ''good'' fit and meaningful parameter error estimates, the error limits in both x and y directions are enlarged by multiplying them with such a number that the fitting procedure yields a reduced chi-square $\chi_{\mathrm{red}}^2 = 1$. Equation \ref{chi} shows that the multiplying factor is equal to the square root of the reduced chi-square of the fit with the original error estimates. This method was used for all of the statistical testing of fit parameters in this paper.

\end{appendix}

\end{document}